\documentclass[article]{aa}
\usepackage{psfig}

\def\l{\label}

\def\beq{\begin{equation}}
\def\bfg{\begin{figure}}
\def\efg{\end{figure}}
\def\eeq{\end{equation}}
\def\bea{\begin{eqnarray}}
\def\eea{\end{eqnarray}}

\def\xs{\{x\}_s}

\def\ltsima{$\; \buildrel < \over \sim \;$}
\def\simlt{\lower.5ex\hbox{\ltsima}}
\def\gtsima{$\; \buildrel > \over \sim \;$}
\def\simgt{\lower.5ex\hbox{\gtsima}}

\begin{document}
\thesaurus{
        (12.12.1;  
Universe
        11.03.1;  
          }

\title{  Detection of non-random patterns in 
cosmological gravitational clustering}  
\author{R.~Valdarnini \inst{1}}

\institute{SISSA, via Beirut 2-4, 34014, Trieste, Italy}

\date{Received \dots; accepted \dots}
\maketitle
\markboth{Nonrandom patterns in gravitational clustering  }{}

\begin{abstract}
A new method for analyzing point patterns produced by the
evolution of gravitational clustering is presented. The method
is taken from the study of molecular liquids, where it has been introduced for
making a statistical description of anisotropic distributions. The statistical 
approach is based on the spherical harmonic expansion of angular correlations.
A general introduction to the method is given; a theoretical
analysis shows that it is partially connected with previous harmonic analyses
applied to galaxy catalogs.
The effectiveness of the statistical analysis in quantifying
clustering morphology is illustrated by applying the statistical 
estimators to point distributions produced by an ensemble of 
cosmological $N-$ body simulations with a CDM spectrum.
The results demonstrate that the statistical method is able to
detect anisotropies in non-random patterns, with different scales 
being probed according to the expansion coefficients.

\end{abstract} 

\section { Introduction}

According to the standard picture the observed structures
in the Universe have arisen via gravitational 
instability from the time evolution of the
initial matter density fluctuation field ${\delta_m(\vec x)}$.
At very early epochs the field is assumed to be a random Gaussian process
(e.g. \cite{pe80}).
In the proposed scenarios  the formation of structures
proceeds hierarchically, with galaxies forming first and larger structures
following later from clustering of galaxies.
The present galaxy distribution shows that clustering is ranging from
small groups up to clusters and superclusters.
A striking feature of the observed large-scale galaxy distribution,
 as revealed by extended ($\simeq 100-200Mpc$) redshift surveys
 (\cite{ge89}; \cite{co94}; \cite{la96}; Broadhurst et al. 1990; 
\cite{sa91}), is that 
 galaxies are connected in a web of sheet-like structures, with 
 voids between them and filaments at the intersections (\cite{ge89}).
 The size of these structures can be as large as $ 50\simeq100 Mpc$.
 These features revealed in the galaxy distribution have been
 formed through gravitational clustering and are connected to the
 initial power spectrum $ P(k)= < |\delta_m(k)|^2>$; thus the observed
 galaxy clustering can in principle be used to put constraints on the 
 allowed cosmological models, for which the fundamental parameters 
 determine the shape of $P(k)$.
 The galaxy distribution today is highly non-linear and $N-$body 
 methods have been used to sample the initial phase space distribution 
 and to study the time evolution of the clustering growth 
 (\cite{ef85}; Bertschinger \& Gelb 1991).
 These methods lead to a final particle distribution which should be
 a representative sample of the expected galaxy distribution
for the cosmological model under study.
The use of $N-$body codes implicitly makes the assumption that galaxies trace 
the matter distribution.
In order to compare the observed galaxy clustering with the results of 
numerical simulations one has to introduce a statistical descriptor.
The method used to investigate the statistical distribution
of galaxies is therefore very important because it should be used 
to discriminate between different models, furthermore the method must also 
provide a mathematical description for the rich variety of clustering 
morphology as seen in the galaxy distribution.

The first method to be introduced as a tool for studying 
galaxy clustering was the 2-point correlation function 
$\xi(r)$ (\cite{to69}; \cite{pe74}). Analyses of galaxy catalogs
show that $\xi(r)$ has a power-law shape $\xi \propto r^{-\gamma}$, 
with $\gamma =1.77$ and is equal to unity for $r_o=(5 \pm 1) h^{-1}Mpc$
\footnote{Here $H_0=100h Km sec^{-1} Mpc^{-1}$ is the present value of the 
Hubble constant} (\cite{pe74}; \cite{fi94}).
The 2-point correlation function has been widely used to constrain cosmological models from results of numerical simulations (\cite{je98} and references
cited therein). 
However the 2-point function does not provide useful information about the 
rich variety of structures which characterize the morphology of the
galaxy distributions.
The reason is that the power spectrum $P(k)$ ( the Fourier transform of
$\xi$ ) does not describe the correlation of the phases that the
$\delta_{\vec k}$'s
develop during the clustering evolution.

For a complete description of the clustering one has in principle to 
consider correlation functions $\xi_N$ of higher order. The measurement
 of these correlations
is impractical however when $N \geq 5$ because of the noise introduced 
by the finite number of
galaxies in the sample (\cite{sa96}). For this reason different 
statistical
tools have been introduced for studying the clustering of the galaxy 
distribution.
These methods are : the void probability distribution $P_0(V)$ (White 1979;
\cite{vo94}; \cite{gh97}); moments of counts in cells (\cite{sb91});
probability distribution of count in cells (\cite{bo93}; \cite{ue96}).

While these methods still make use of the higher order moments of the
clustering distribution, alternative approaches have been introduced 
which are more geometrical and implicitly contain information from 
the $N-$ point correlations at all orders.
The first method to be introduced was the percolation statistic
(\cite{ze82}; \cite{kl93}), other methods include the topological
genus-density threshold (\cite{go86}; Ryden et al. 1989), minimal spanning 
trees (Barrow, Bhavsar \& Sonoda 1985);
Minkowski functionals 
(Mecke, Buchert \& Wagner 1994; \cite{ke97}), graph-theory (\cite{ue99})
or a global descriptor
(J-function) based on the nearest-neighbor distribution (\cite{ke99}).
In order to quantify the geometrical features of the galaxy 
clustering several statistics have been explicitly developed to single out
in a quantitative way the filaments and sheets of the cosmic network.
These statistics are based on the moments of the mass distribution 
and can be used to detect filaments ({\cite{fr86}; \cite{da99}; \cite{ba92}),
 or the
shape of the clustering (\cite{lu95}; \cite{ro96}). The shape 
statistic can also be constructed using Minkowski functionals 
(\cite{sc97}; \cite{bh20}).

The existence of this variety of approaches is
related to the fact that there is not any single well defined theoretical 
method which can be used to analyze the shape of the clustering network.
Furthermore a statistical method must also be able to produce a robust
statistical measure, in order to be applied to galaxy catalogs.
These requirements are naturally satisfied by the 2-point function $\xi$,
but, as outlined above, this function it is not indicated for 
detecting filaments.

 In this paper, I propose an alternative method for analyzing the clustering
 morphology which is a generalization of the 2-point function $\xi$ and  
 is based on the spherical harmonic analysis.
  The statistical method has been applied in molecular fluid dynamic 
  simulations for describing the structures of anisotropic fluids 
 or disordered metallic glasses (\cite{st83}; \cite{wa91}).
The method is applied here for illustrative purposes to point distributions
obtained from cosmological $N-$body simulations.

The paper is organized as follows: In Sect.2, I present the method. Sect.3 
 describes the ensemble of CDM cosmological simulations  
used to test the effectiveness of the method.
In Sect.4  the statistic is applied to the point sets produced by 
the simulations, the main results are discussed and the conclusions 
are summarized.
\section{ THE METHOD}
\subsection{ Spherical harmonic analysis}
For describing galaxy clustering evolution in the non-linear regime,
 use is made of different statistical methods, based on the galaxy positions,
 which have been outlined in the Introduction. The approach
introduced here makes use of the positions  of the point 
distribution within a certain distance from a randomly chosen one.

The method is drawn from molecular dynamics simulations,  where it 
has been introduced 
for studying orientational order of supercooled liquids and
metallic glasses. A general review can be found in 
 Haile \& Gray (1980) and Mc Donald (1986),
here I will follow the notation of Wang \& Stroud (1991).

Let us consider a system of $N_p$ particles.
  The $i-th$ particle has coordinates $\vec r_i$, in 
  an arbitrary reference frame.
For a specified cutoff radius $R_c$ all the particles such that 
 $|\vec r_i-\vec r_j|< R_c$ are neighbors of $i$.

The line joining $i$ to one of the $j$ is termed a {\it bond}.

The angular coordinates of the vector $\vec \Delta_{ji}\equiv \vec r_j - \vec r_i $
are $\theta_j, \phi_j$ and the quantity 
\smallskip
\beq
Q_{lm}(\vec r_i)=\sum_{j \ne i} Y_{lm}(\theta_j,\phi_j),
\l{eq:defQ}
\eeq
is the coefficient of the spherical harmonic expansion of the angular density
of the bonds associated with the particle $i$.

In Eq. \ref{eq:defQ}, and hereafter, summation is understood over all
 particles $j$ of the distribution such that 
 $|\vec r_i-\vec r_j|< R_c$.

The coefficients $Q_{lm}(\vec r_i)$ are defined as the 
bond-orientational order parameters (\cite{wa91}) and they
can be drastically changed by a rotation of the reference systems.
A natural quantity to consider , which is rotation invariant, is
\beq
Q_l(\vec r_i)=\sqrt {{{4\pi}\over{2l+1}} \sum_{m=-l}^{m=l}
Q^{\star}_{lm}(\vec r_i) Q_{lm}(\vec r_i)}.
\l{eq:Qla}
\eeq

Using the addition theorem for the spherical harmonics, the expression 
simplifies to

\beq
Q_l(\vec r_i)=\sqrt { \sum_j \sum_k P_l(\gamma_{jk})},
\l{eq:Qlb}
\eeq
where $P_l$ is the Legendre polynomial, 
$$
\gamma_{jk}\equiv cos(\theta_{jk})=\vec \Delta_{ji}\cdot\vec\Delta_{ki}/
(|\vec \Delta_{ji}||\vec \Delta_{ki}|)\nonumber
$$
is the angle between two bonds and  the summations in  Eq.
\ref{eq:Qlb} are then independent of the chosen frame.

The expression  for $Q_l(\vec r_i)$ can be further simplified 
if one considers that $\gamma_{jj}=1$, furthermore it is more convenient
to redefine $Q_l(\vec r_i)$ dividing by the
mean number of neighbors $<N_i>=\bar N_i$ of the particle $i$ :

\beq
Q_l(\vec r_i)={{1} \over {\bar N_i}} \sqrt { N_i +2 \sum_j \sum_{k>j} P_l(\gamma_{jk})}.
\l{eq:Qlc}
\eeq

For a random distribution the summation terms in the square root 
are negligible for $N_i \gg 1$,
thus $Q_l(\vec r_i)\propto 1 / \sqrt {N_i}$ for Poisson noise.

For the whole system an order parameter $<Q_l>$ can be defined by
averaging over all of the $N_p$ particles:
\beq
<Q_l>=\frac {1}{N_p} \sum_{i=1}^{N_p} Q_l(\vec r_i) \equiv Q_l.
\l{eq:Qld}
\eeq

Thus the clustering distribution can be studied by evaluating
the $Q_l$ at a specified cutoff radius. 
Going to higher coefficients ($ l \gg 1$) means probing 
smaller scales in the bond distribution.

However the $Q_l$ do not exhibit spatial information, and
a more useful quantity is the auto-correlation 
   $ G_l(r)$  of the coefficients
$Q_l(\vec r_i)$. The function $G_l$ is defined as follows:
for all of the $M_p$ pairs $(i,k)$, such that 
$|\vec r_i-\vec r_k|=r \pm \Delta r$, where $\Delta r$ is the 
thickness of the radial bin, then
  $G_l(r) $ is the sum over all of these pairs 

\beq
G_l(r)={{1}\over{M_p}} \sum_i \sum_k{{4\pi}\over{2l+1}} 
\sum_{m=-l}^{m=l} Q^{\star}_{lm}(\vec r_i) Q_{lm}(\vec r_i+\vec r).
\l{eq:GL1}
\eeq

This equation can be greatly simplified: let
 ${j}$ be the set of neighbors of the particle $i$ 
and  ${p}$ that of the particle $k$, which satisfy
 $|\vec r_j-\vec r_i|< R_c$  and $|\vec r_p-\vec r_k|< R_c$. 
Then the summation becomes
\begin{eqnarray*}
\sum_{m=-l}^{m=l} Q^{\star}_{lm}(\vec r_i) Q_{lm}(\vec r_i+\vec r) = 
\sum_{m=-l}^{m=l} Q^{\star}_{lm}(\vec r_i) Q_{lm}(\vec r_k)= \\
\sum_{m=-l}^{m=l} \sum_j Y^{\star}_{lm}(\theta_j,\phi_j) 
\sum_p Y_{lm}(\theta_p,\phi_p).  
\end{eqnarray*}
\smallskip
The final expression for $G_l$ is:
\beq
G_l(r)={{1}\over{M_p}} \sum_i \sum_k \sum_j \sum_p P_l(\Gamma_{jp}),
\l{eq:GL4}
\eeq
with {\bf $\Gamma_{jp}$} being the angle between $\vec \Delta_{ji}$ and 
$\vec \Delta_{pk}$.  The summation over the pairs  is   
 $\sum_i \sum_k$, with the sum over the particles $k$ 
only for those particles with $|\vec r_i-\vec r_k|=r \pm \Delta r$.  

Note that now it is not possible to simplify the equation 
as was done for Eq. \ref{eq:Qlc}, because one is considering 
the correlation between 
{\it different} order parameters.

The information about the degree of angular correlation between bonds
contained in the $G_l(r)$ is more explicit if the $G_l$ 
themselves are divided by  the zero-order correlation function
\beq
G_0(r)={{4\pi}\over{M_p}} \sum_i \sum_k
Q^{\star}_{0}(\vec r_i) Q_{0}(\vec r_i+\vec r).
\l{eq:GL2}
\eeq
The functions $G_l(r)$ can be made dependent also 
on the direction of $\vec r$ by considering all the pairs with 
orientation of the relative separation $\vec r$ in the solid 
angle range $\Omega_{\vec r} \pm \Delta \Omega_{\vec r}$.
These functions $G_l(\vec r)$ are of scarce practical use because 
of the noise level which is introduced when applied to a finite sample.


\subsection { Theory}
The equations that have been obtained for $Q_l$ and $G_l(r)$ are
defined through a point set distribution; from  a theoretical point of view 
it is possible to relate these functions to the power spectrum of the
density fluctuations. Spherical harmonic analysis has been widely applied
in cosmology, both to the galaxy distribution (\cite{sc92}; \cite{sc93};
\cite{ba95}) and to the study of
cosmic background radiation maps (\cite{hu97}; \cite{ba99}). 
I will follow here the notation of Scharf et al. (1992), who have analyzed the angular
distribution of the IRAS redshift survey using spherical harmonics.
For the generic distribution considered in Sect.2.1
let us define  $n(\vec r)$ to be the number density and
 $\bar n$ to be the average one. It is also useful to introduce
 $\Phi(r)$ as a generic selection function, 
 here $\Phi=\bar n$.

Then the coefficients $Q_l^m(\vec r_o)$, $\vec r_o$ being the center position,
are given by
$$
Q_l^m(\vec r_o)=\int \Phi(\vec r^{\prime}) {{n(\vec r^{'})} \over {\bar n}} 
 Y_l^m(\Omega_{\vec r^{'}}) d^3 r^{'} ~~~~\vec r=\vec r_o+\vec r^{\prime},
$$
 where the integral is taken over a sphere of radius $R_c$.
The number density is
$$ 
n(\vec r)=\bar n [ 1+ \delta(\vec r)]=\bar n[1+\delta(\vec r_o+\vec r^{'})],
$$
here $\delta(\vec r)$ is the density fluctuation.
Thus the equation for $Q_l^m(\vec r_o)$ becomes 
\bea
Q_l^m(\vec r_o)& =& \int \Phi [1+\delta] Y_l^m d^3 r^{'}= \nonumber \\
& & \int \Phi Y_l^m d^3 r^{\prime}+\int \Phi(\vec r^{'}) \delta 
 (\vec r_o+\vec r^{'}) Y_l^m d^3 r^{'}, \nonumber
\eea
where the first term on the rhs is zero for $l \ge 1$. The density fluctuation
$\delta(\vec r)$ can be expanded into Fourier modes, and the final expression
for $Q_l^m(\vec r_o)$  is

\beq
Q_l^m(\vec r_o)=\sum_{\vec k} \delta_{\vec k} e ^{ i\vec k \vec r_o}
\int \Phi(\vec r^{'}) e^{i \vec k\cdot r^{'}}  Y_l^m(\Omega_{\vec r^{'}}) d^3 r^{'}.
\eeq
The $ e^{i\vec k \vec r }$ is now decomposed into spherical functions,
so that the ensemble average gives
\begin{eqnarray}
< |Q_l^m|^2 >& = &( 4\pi)^2 \sum_{\vec k} |\delta_{k}|^2
 \left( \int_0^{R_c} \Phi(x) x^2 j_l(kx) dx \right) ^2 \nonumber \\
& & |Y_l^m(\Omega_{\vec k})|^2,
\l{eq:qlm}
\end{eqnarray}
 here the $j_l(x)$s are the spherical Bessel functions.
The summation over $k$ can be translated into an integral  and
$\int |Y_l^m|^2 d\Omega=1$, so that 
\bea
< |Q_l^m|^2 >& =& V {{2} \over {\pi}} \int_0^{\infty} P(k)k^2 dk 
 \left[ \int_0^{R_c} \Phi(x) x^2 j_l(kx) dx \right] ^2 \nonumber \\
& & + {\cal N},
\l{eq:QF}
\eea
where $P(k)=<|\delta_k|^2>$, $V$ is a normalizing volume and
$\cal N$ is the shot noise term ${\cal N} =\int \Phi(x)x^2 dx$. 

The coefficients $C_l$ are defined as
\beq
C_l^2={ {1} \over { {\cal N} (\Omega/4\pi)(2l+1)}} \sum_m |Q_l^m|^2,
\l{eq:Cla}
\eeq
and assuming statistical isotropy they 
can be related to Eq.(\ref{eq:Qld}) by
$Q_l\simeq C_l/ {\bar N_i^{1/2}}$.
For an uncorrelated distribution $C_l=1$, furthermore the integral
 $ \int_0^{R_c} \Phi(x) x^2 j_l(kx) dx \equiv\Psi_l(kR_c)$ defines
 a bidimensional window function in the parameter space $(R_c,l)$.
Increasing the value of $l$ means probing smaller angular scales,
the value of  the cutoff radius $R_c$ sets the size of the spectrum
contribution to the integral (see,  for example, Fig. 4 of \cite{sc93}).
The statistical method introduced in Sect.IIA, and
originally proposed in a different field, is then mathematically equivalent
to the spherical harmonic analysis which has been applied to
galaxy surveys (\cite{sc93}).
The order parameter $Q_l$ is just related to the angular power spectrum
estimator $C_l$ (\cite{sc92}).

However Eq. (\ref{eq:qlm}) is valid as long as the position of the observer  
located at $\vec r_0$ can be considered as random with respect to the bond
distribution inside the sphere of radius $R_c$. This is a good approximation
for values of the cut-off radius  $R_c $ much larger than the
2-point clustering length, a condition which is valid for the analyzed 
angular catalogs.  In Sect. 4 the statistical method is tested 
using the point distributions obtained from a set of cosmological 
simulations, and the coefficients $Q_l$ are calculated for different 
parameters and cut-off radii $R_c$. 
Therefore one has to consider in principle also the three-point 
correlation terms  between the particle located at $\vec r_0$ and the \\ other 
particles in the bond sphere. The ensemble average for $ |Q_l^m|^2 $ is 
then (Peebles 1980, sect. 36B):

\bea
< |Q_l^m|^2 > & = &\int \Phi d \vec x  \int \Phi d \vec y [ 1+\xi(x) + 
\xi(y) +\xi(|\vec x - \vec y|) \nonumber \\
 &  & + \zeta(x,y,|\vec x - \vec y|) ] Y_l^m(\Omega_{\vec x}) 
{Y^{\star}_l}^m(\Omega_{\vec y}), 
\l{eq:Qg3}
\eea
where $\vec x$ and $\vec y$ are the particle separations relative to the origin 
  and $\zeta$ is the reduced three-point correlation 
function. Integration over solid angles removes for $l\geq1$ the first three 
terms on 
the rhs of the equation. Of the remaining terms, the first   
( $ \xi(|\vec x - \vec y|= \sum_{\vec k}  |\delta_k|^2 e^{i \vec k( \vec x - 
\vec y)})$,
 gives Eq. (\ref{eq:qlm}), while the other 
takes into account for the triangle configuration the  correlations between
the three particles. The integral can 
be evaluated in terms of the Fourier transform of $\zeta$ (\cite{fr84}), the
 bispectrum $B(\vec k_1, \vec k_2, \vec k_3)$ :

\bea
\int \Phi d \vec x  \int \Phi d \vec y \zeta(x,y,|\vec x - \vec y|)Y_l^m(\Omega_{\vec x}) {Y^{\star}_l}^m(\Omega_{\vec y}) =  & & \nonumber \\  
\sum_{\vec k_1 \vec k_2 \vec k_3} B(\vec k_1, \vec k_2, \vec k_3)
\int \Phi d \vec x  \int \Phi d \vec y Y_l^m(\Omega_{\vec x}) 
{Y^{\star}_l}^m(\Omega_{\vec y}) & & \nonumber \\  
e^{i(\vec k_1 \vec x_1+ \vec k_2 \vec x_2 + \vec k_3 \vec x_3)}=   
\sum_{ \vec k_2 \vec k_3} B(\vec k_1=-(\vec k_2+\vec k_3), \vec k_2, 
\vec k_3) &  &\nonumber \\
\int \Phi d \vec x  \int \Phi d \vec y Y_l^m(\Omega_{\vec x}) 
{Y^{\star}_l}^m(\Omega_{\vec y})   
e^{i( \vec k_2 \vec x + \vec k_3 \vec y)} =  & & \nonumber \\
 (4 \pi )^2  (-)^l 
 \sum_{ \vec k_2 \vec k_3} \int \int B(\vec k_1, \vec k_2, \vec k_3) 
  \Psi_l(k_1R_c)   \Psi_l(k_2R_c) & & \nonumber \\
 Y_l^m(\Omega_{\vec k_1}) {Y^{\star}_l}^m(\Omega_{\vec k_2})~, ~~~~~~~~~~~~~~~~~~~~~~~~~& &
\l{eq:bs}
\eea
where the constraint $\sum \vec k_i=0$ is required by homogeneity, 
$\vec x \equiv \vec x_2 - \vec x_1$ and $\vec y\equiv \vec x_3 -\vec x_1$.
In second order perturbation theory, for Gaussian initial conditions, 
the bispectrum can be expressed as ( Fry 1984 ; \cite{ve20}) :
\beq
 B(\vec k_1, \vec k_2, \vec k_3)= K( \vec k_1, \vec k_2) P(k1) P(k2) + 
cyc.~, 
\l{eq:bs1}
\eeq
here $cyc.$ means cyclic terms.
The term $K(\vec k_1, \vec k_2)$ is weakly dependent on the cosmology and 
can be written as (Fry 1984; Goroff et al. 1986; \cite{ve20}):
\beq
 K( \vec k_1, \vec k_2)= A_0+A_1 cos(\theta_{12})+A_2 cos^2(\theta_{12}),
\l{eq:bs2}
\eeq
where the coefficients $A_i$ depend on the assumed biasing expansion 
for the local density field and $\theta_{12}$ 
is the angle between $\vec k_1$ and $\vec k_2$.
The integral (\ref{eq:bs}) can then be evaluated for a particular cosmological
model. The dependence of $B$ on the triangle shape in the $k$-space shows
that the integral is non zero only for $l \leq2$. The integral (\ref{eq:bs}) 
is a particular case of the integrals considered by Verde, Heavens and 
Matarrese (2000). These authors have expanded the projected galaxy
density in spherical harmonics and studied the 3-point function of the 
expansion coefficients, which is a quantity directly related to the 
bispectrum.
The evaluation of the above integral is rather complicated (see, e.g., Eq. (28) 
of Verde, Heavens and Matarrese 2000) and will not be considered here.
In Sect. 4 it will be seen that the corrections (\ref{eq:bs}) 
to Eq. (\ref{eq:qlm}) must be rather small already for $R_c \simeq 2 r_0$.

This theoretical analysis can also be used to derive an
 expression for the quantity $G_l(\vec r)$, which contains
spatial information about the bond angular distribution.
The evaluation of $G_l(\vec r)$ now involves correlation terms up to the
fourth order. In analogy with Eq. (\ref{eq:Qg3}) 
one has to compute for  $G_l(\vec r)$ the ensemble average 
\bea
\lefteqn{ G_l(\vec r)=<Q_l^m(\vec r_o) Q_l^{\star m}(\vec r_o+\vec r) > = } \nonumber \\
\,& \int \Phi d \vec x  \int \Phi d \vec y [ \xi(|\vec r+ \vec y - \vec x|)+
\zeta(x,|\vec r + \vec y - \vec x |, |\vec r + \vec y|) \nonumber \\
\,&+\zeta(|\vec r + \vec y - \vec x |, y, |\vec r - \vec x|) +  
\xi(|\vec r + \vec y|) \xi(|\vec x -\vec r|)+ \xi(r)\nonumber \\
\,&\xi(|\vec r + \vec y - \vec x)|)+ 
\eta (x,r,y,|\vec r + \vec y|,|\vec r - \vec x|, |\vec r + \vec y - \vec x|)]\nonumber \\
\,&{ Y_l^m(\Omega_{\vec x}) {Y^{\star}_l}^m(\Omega_{\vec y}),~~~~~~~~~~~~
~~~~~~~~~~~~~~~~~~~~~~~~~~~~~~~~~} \l{eq:Ga1}
\eea
where $\eta$ is the reduced four-point correlation function, and
 only the terms which are non-zero after the integration 
over the solid angles are shown in the integrals;
the separations of the four-particle configuration of Eq. (\ref{eq:GL4})
are defined as:~$ \vec r= \vec r_k - \vec r_i  ,~ \vec y= \vec r_p - \vec r_k ,
~ \vec x = \vec r_j - \vec r_i$. The integrals in (\ref{eq:Ga1}) can be 
evaluated using the same method adopted  for the integral (\ref{eq:bs}).
As previously discussed, from the results of Sect. 4 for $< |Q_l^m|^2 >$ ,
the third and fourth-order terms are expected to be negligible.
I will consider here only the leading order term 
$ \xi(|\vec r+ \vec y - \vec x|)$ and the ensemble average
for $G_l(\vec r)$ is then

\bea
\lefteqn{ <Q_l^m(\vec r_o) Q_l^{\star m}(\vec r_o+\vec r) > = } \nonumber \\
\,& \sum_{\vec k} e^{-i \vec k \vec r} |Y_l^m(\Omega_k)|^2 \left [
 \int \Phi j_l(kx)x^2 dx \right]^2 (4\pi)^2 |\delta_k|^2. \l{eq:Ga}
\eea

The summation over $\vec k$ becomes an integral and 
if $e^{-i \vec k \vec r}$ is now decomposed as
\beq
e^{-i \vec k \vec r} = 4 \pi \sum_{pq} i^p j_p(kr) Y^*_{pq}(\Omega_{\vec r})Y_{pq}(\Omega_{\vec k}),
\eeq
then the triple integral for the spherical harmonics $Y_l^m$ in 
Eq.(\ref{eq:Ga}) is 
\begin{eqnarray*}
\lefteqn {\int Y_{pq} Y_{lm}^* Y_{lm} d\Omega_k=\int Y_{pq}(-)^mY_{l-m}Y_{lm}d\Omega_k= } \nonumber \\
\,& (-)^m \left [ {(2l+1)}^2~({2p+1)} \over {4\pi} \right ]^{1/2}
\left ( \begin{array}{lcr}
l & l & p \\
0 & 0 & 0 
\end{array} \right )   
\left ( \begin{array}{lcr}
l & l & p \\
m & -m & q 
\end{array} \right ).
\end{eqnarray*}
The symbols
$ \left ( \begin{array}{lcr}
l_1 & l_2 & l_3 \\
m_1 & m_2 & m_3 
\end{array} \right )   
$
are the Wigner $3-j$ coefficients. Because of their properties, the 
summation over the $q$ terms in 
Eq. (\ref{eq:Ga}) is non-zero only for $q=m-m=0$, furthermore the
$p$ summation admits only terms with $2l+p=even, 0 \leq p \leq 2l$.
Finally the expression for $G_l(\vec r)$ becomes :
\bea
  G_l(\vec r)  =  { {4\pi}\over {2l+1}} \sum_m
 <Q_l^m(\vec r_o) Q_l^{\star m}(\vec r_o+\vec r) >=  \nonumber  \\
 {{(4\pi)^4}\over {(2\pi)^3}} V \int_0^{\infty} dk k^2 
\left [ \int_0^{R_c} \Phi x^2 j_l(kx) dx \right]^2 P(k) \nonumber  \\
\sum_{p( even )=0}^{2l} i^p j_p(kr) Y_{p0}^{\star}(\Omega_{\vec r})
\sum_{m=-l}^{m=l} (-)^m {{(2p+1)}^{1/2} \over {(4\pi)^{1/2}}} \nonumber \\
\left ( \begin{array}{lcr}
l & l & p \\
0 & 0 & 0 
\end{array} \right )   
\left ( \begin{array}{lcr}
l & l & p \\
m & -m & 0 
\end{array} \right ).
\l{eq:Gb}
\eea
In the above equation  the function $G_l$ still depends on the 
orientation of $\vec r$ through the $m$ terms. However, owing to 
statistical isotropy, this dependence should not be present.
In fact, it can be shown that (\cite{ga95})
$
\sum_m (-)^m 
\left ( \begin{array}{lcr}
l & l & p \\
m & -m & 0 
\end{array} \right )   
=(2l+1)^{(1/2)}(-)^l \delta_{p,0},
$
and Eq. (\ref{eq:Gb}) simplifies to

\bea
\lefteqn { G_l(r) ={ {(4\pi)}^3 \over {(2\pi)^3}} V \int_0^{\infty} dk P(k) k^2 } \nonumber \\
\,& \left[ \int_0^{R_c} \Phi(x) x^2 j_l(kx) dx \right] ^2
  j_0(kr).
\l{eq:Gc}
\eea

From this function one can recover the standard definition for $\xi(r)$
taking for $l=0$ the limit $ R_c \rightarrow 0 , 
\int_0^{R_c} \Phi(x) x^2 j_0(kx) dx \rightarrow 1/4\pi$.
The statistical descriptor of clustering defined by Eqs. (\ref{eq:Qld}) 
and (\ref{eq:GL4}), 
when applied to a spatial
point process, should be able to describe in a quantitative way 
the patterns generated by the clustering distribution.
As outlined in the Introduction different methods have been employed
to this end; the most important advantage of the order parameters 
introduced here is that they can can be used to study spatial
patterns in a well-defined way by varying the angular scale $l$ or
the depth $R_c$ of the bond spheres.
In order to test the clustering features described by the order
parameter $Q_l$ and correlation function $G_l$ , I have then 
analyzed point distributions generated by a set of cosmological
$N$-body simulations with CDM spectra.
The simulations are evolved in time starting from an early epoch when
linear theory applies. 
The clustering evolution gives rise  to a variety of spatial
 structures and the statistical descriptors are then applied, 
in order to analyze the produced patterns,
to numerical outputs selected at different epochs. 
The results obtained are discussed in Section 4.

\section{ THE SIMULATIONS}
The point distributions used for the statistical analysis have been 
constructed from a set of purely gravitational 
cosmological $N$-body simulations.
For each of these simulations
the clustering evolution of $N_p=10^6$ particles is followed in time 
 in a box of comoving size $L=200 h^{-1}Mpc$.
The gravitational potential is solved using a $P^3M$ code (\cite{ef85}), with
$N_m=256^3$ grid points. The interparticle force has a linear  
shape cloud profile with a comoving softening parameter  $\varepsilon=0.220Mpc$.

The initial conditions for 
the particle positions and velocities have been generated at the 
initial time $t_i$ according to the standard Zel'dovich approximation (\cite{ef85}).
The power spectrum $P(k)$ is that of a standard CDM model (\cite{ba86}) 
with $\Omega_m=1$ and $h=0.5$.
The Fourier components $\delta_k$ are Gaussian distributed, with 
random phases and variance $P(k)$.
An ensemble of five different integrations has been carried out,
each with a different random realization for $\delta_k$.
The simulations are integrated in time with the  expansion factor $a(t)$ 
being used as the integration variable and with $a(t_i)=1$.
Records of particle positions have been taken at $a(t)=3.0, 3.8, 4.5,6,7.9$.
The power spectrum is normalized so that the rms mass fluctuation 
in a $R=8h^{-1}Mpc$ sphere takes the value $\sigma_8=0.4$ at $a(t)=3.0$.

In order to measure the order parameters associated with peaks above a
given height of the underlying density field $\delta(\vec x)$ I have used 
the peak-background formalism (\cite{wh87}).
For each particle $i$ the peak number $n_i$ is calculated at $a(t_i)$ from
the initial density field; $n_i$ is the number of peaks 
with height $ \delta_s \geq \nu \sigma_s$, here $\delta_s$ is the field
$\delta(\vec x)$ Gaussian smoothed with $R_s=1.1 Mpc$ and $\sigma_s^2$ is
the variance of $\delta_s$.
The field is constrained to take the value $\nu_b \delta_b$ when
smoothed on a scale $R_b > R_s$ (\cite{wh87}; \cite{pa91})
The particle $i$ is then identified as a 'peak' if $n_i >p$, with $p$ being a 
uniform random number between $0$ and $1$.

The total number of particles associated with the peaks of a given 
 height is $\simeq \sum n_i $.
Three different threshold levels have been chosen:
$\nu=0.5,1.3,2.5$ which correspond, respectively,
to $\sum n_i \simeq 250,000,~ 165,000,~ 33,000$ in the simulation box.
Hereafter the notation $\{x\}_s$ means the subset of the $10^6$ simulation
particles selected to identify the peaks with threshold $\nu=s$.
At different epochs $a(t)$ the order parameters have been computed 
by applying the statistical estimators
to the point distribution produced by $\{ x\}_s$.
The clustering of the mass distribution produced 
by all of the simulation particles has been also considered, 
 in this case $\delta \geq s=-1$.

\begin{figure} 
\centerline{
\psfig{file=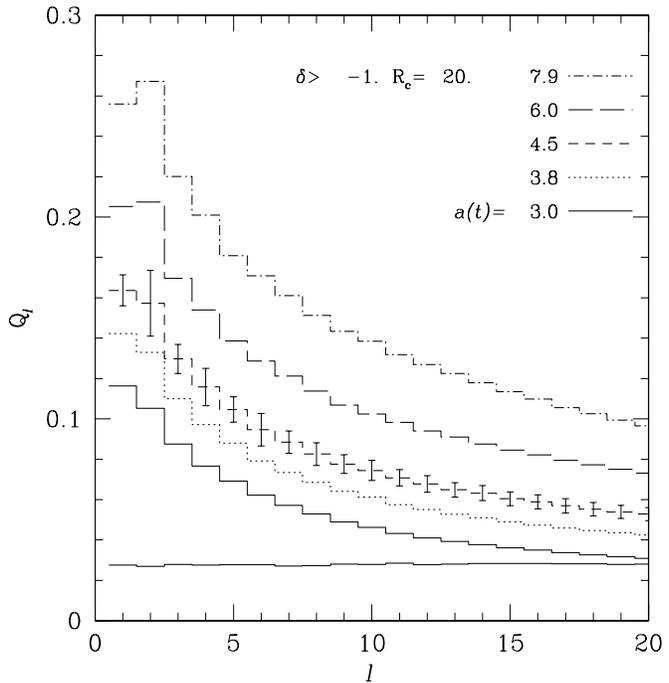,height=4.in,width=3.8in}
}
\caption[]{ The coefficients $Q_l$ are plotted as functions of {\it l}
for different expansion factors $a(t)$. 
The estimator (\ref{eq:Qld}) has been applied 
to the whole point distribution of the simulation particles.
Here and in the following plots the cut-off radius $R_c$ is in 
units of $h^{-1}Mpc$.
Error bars show the scatter within the five simulation ensemble.
The continuous line with a nearly constant value for the $Q_l$ refers to a 
random test distribution. 
}
\label{Qa}
\end{figure}

\section{ RESULTS AND CONCLUSIONS}
The order parameters $Q_l$ are shown in Figs. \ref{Qa},~\ref{Qb} \& \ref{Qc}
for different values of the expansion factor $a(t)$, threshold levels 
and cut-off 
radius  $R_c$. In Fig. \ref{Qa} the time evolution of $Q_l$ is plotted 
as a function of $l$ for different epochs.
Several generic features can be seen in the plot: the coefficients $Q_l$ 
grow with time as the clustering develops and decay with increasing $l$ as 
smaller angular scales 
are probed. This follows directly form the shape of $P(k)$.
 For a random test distribution  with $N_i \simeq 10^3$ neighbors 
 $Q_l \simeq const \simeq 1 / \sqrt {N_i}$,  
and the estimator 
(\ref{eq:Qld}) is clearly able to detect non-random distributions
with a high level of significance.
The scatter in the numerical ensemble  can be judged from the size of 
the error bars, for the sake of clarity they have been shown in each plot
only for a single case. For the plot of Fig. \ref{Qa} the point distribution
is that obtained from the $\{x\}_{-1}$ set. In this case the computation of
the $Q_l$ can be greatly reduced by using a random sub-sample of the 
particle set. This has been done also for other sets $\{x\}_{s}$, 
 according to required computational load needed to evaluate the
coefficients $Q_l$. 

\begin{figure} 
\centerline{
\psfig{file=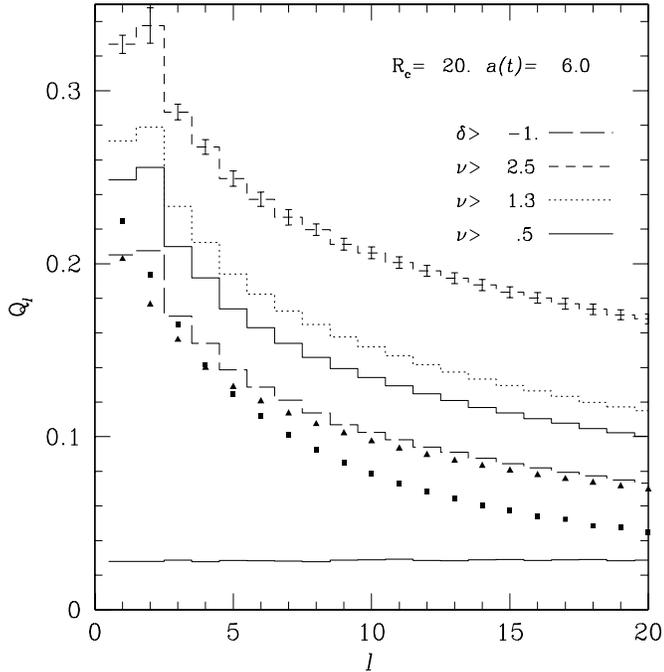,height=4.in,width=3.8in}
}
\caption{
Here the estimators $Q_l$ have been computed at a fixed time and cut-off
radius $R_c$ from the point distributions generated by $\{x\}_{s}$, with
$s=-1,0.5,1.3,2.5$. The black squares are the rescaled linear theory 
coefficients
 $C_l$, computed from the CDM power spectrum at $a(t)=6$ and for $R_c=20$.
The power spectrum has been normalized as in the simulations, so that 
the points compare directly with the $\delta=-1$ $Q_l$ coefficients.
 The black triangles refer to the same theoretical coefficients, but 
 computed in this case using the non-linear power spectrum of the simulation.}

\label{Qb}
\end{figure}

The results obtained have been found to be quite 
robust to changes in the size of the random sub-sample, usually an 
accurate evaluation of $Q_l$ can be obtained already with a $ \simeq
20 \% $ random sub-sample of the particle set.
How the $Q_l$ change when particle populations with different clustering
properties are selected is shown in Fig. \ref{Qb}. For a fixed $a(t)$ and 
$R_c$ the coefficients $Q_l$ have been evaluated for particle subsets 
associated with different thresholds. For any $l$ the coefficients $Q_l$ get
 larger when higher thresholds $\nu$ are selected.
The choice of higher $\nu's$ corresponds to a more clustered distribution,
and this is clearly detected by the coefficients $Q_l$.
It is worth stressing that the trend is obtained according to the adopted
definition (\ref{eq:Qlc}) for $Q_l(\vec r_i)$. The angular power spectrum
coefficients $C_l$ scale approximately as $\propto N_i$, thus the 
enhanced clustering of particle subsets associated with higher $\nu's$ is
 masked for the $C_l$ by the drop in $N_i$, with respect lower thresholds.
The black squares in Fig. \ref{Qb} are the rescaled linear theory 
coefficients (\ref{eq:Cla}). In this case the comparison has been made for
the population $\delta \geq -1$, in order to avoid errors introduced by a
linear bias approximation for Eq. (\ref{eq:Cla}), if subsets $\{x\}_s$ 
corresponding to higher thresholds $\nu$ would have been selected.

\begin{figure} 
\centerline{
\psfig{file=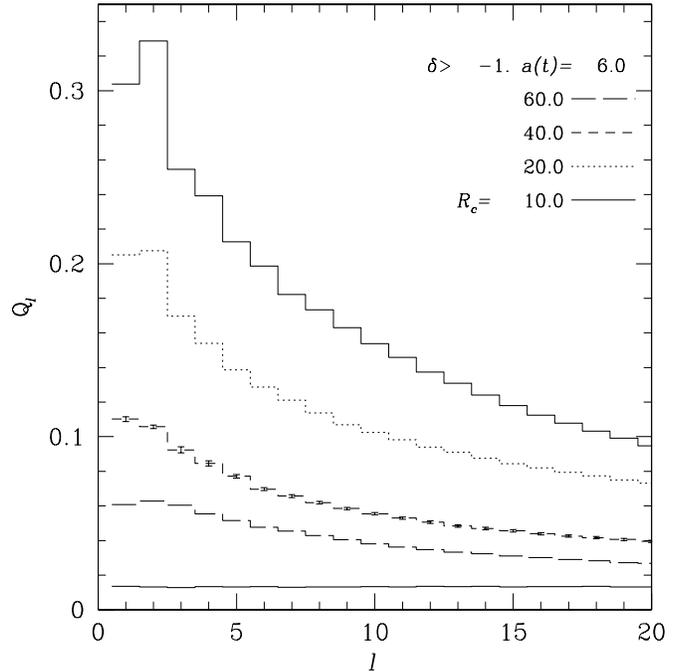,height=4.in,width=3.8in}
}
\caption{ The coefficients $Q_l$ are shown at $a(t)=6$ for four 
different cut-off radii 
$R_c=10,20,40,60$ ( in $h^{-1}Mpc$ units).
The estimators have been applied, as in Fig. \ref{Qa}, to the whole
particle distribution.}
\label{Qc}
\end{figure}

There is a good agreement with the numerical experiment up to $l \simeq 5$;
beyond this angular scale ( $\simeq \pi/5$) non-linearity effects make the
$Q_l$ depart from the linearized coefficients. For the adopted
normalization the size of the non-linear scale is about $R_{NL} \simeq
5-8 h^{-1} Mpc$, at the epoch shown in the plot, which is in rough agreement
with $R_c \Delta \theta$.
When the power spectrum of the simulation particles is used to 
compute the theoretical coefficients there is an excellent agreement with
the corresponding measured $Q_l$ down to the smallest angular scale 
(black triangles of Fig. \ref{Qb}). For $ l \leq 2$ the small differences
show that the corrections (\ref{eq:bs}) to Eq. (\ref{eq:qlm}), for the 
considered cosmological model are negligible
already for $ R_c \simgt 20h^{-1}Mpc$.

The coefficients $Q_l$ are plotted for different cut-off radius $R_c$, in
 Fig. \ref{Qc}, for a fixed time and a given set $\{x\}_s$. There is a 
general tendency for the $Q_l$ to get smaller at a fixed $l$ when
 $R_c$ is increased. This is expected because the bond distribution 
approaches  isotropy as $R_c$ gets higher. This trend is confirmed for all
 values of $l$ and choices of different distributions $\{x\}_s$. Note also
how the scattering in ensemble is reduced ($R_c=40 h^{-1}Mpc$) with 
respect to the smaller values of $R_c$ shown in the previous plots.
The analysis performed so far shows that the order parameters $Q_l$ can be
used as reliable statistical indicators of the global clustering 
properties of a point distribution.

From the histograms of Fig. \ref{Qc} an interesting result is that the
quadrupole coefficient $Q_2$ of the bond angular distribution is larger than 
the dipole coefficient $Q_1$. This is valid with a high significance 
only for values of the cut-off radii $R_c \simlt 2 r_0$, close the value
of the 2-point clustering length. This excess of local anisotropies,
compared to what is found for $R_c \simgt 20 h^{-1} Mpc$, suggests that 
a significant amount of substructure is present in the particle
distribution of dark matter halos.
Jing (2000) has analyzed the halo dark matter 
density profiles from a set of cosmological $N-$body simulations. 
For those halos which are substructure rich and less virialized, 
he found large deviations from the analytical profile of  Navarro, Frenk
 \& White (1995). 
The above results therefore suggest that the proposed statistical method
can be profitably used to correlate in a quantitative way the local
halo anisotropies with other halo properties, when investigating their
evolution in different cosmological models.

\begin{figure} 
\centerline{
\psfig{file=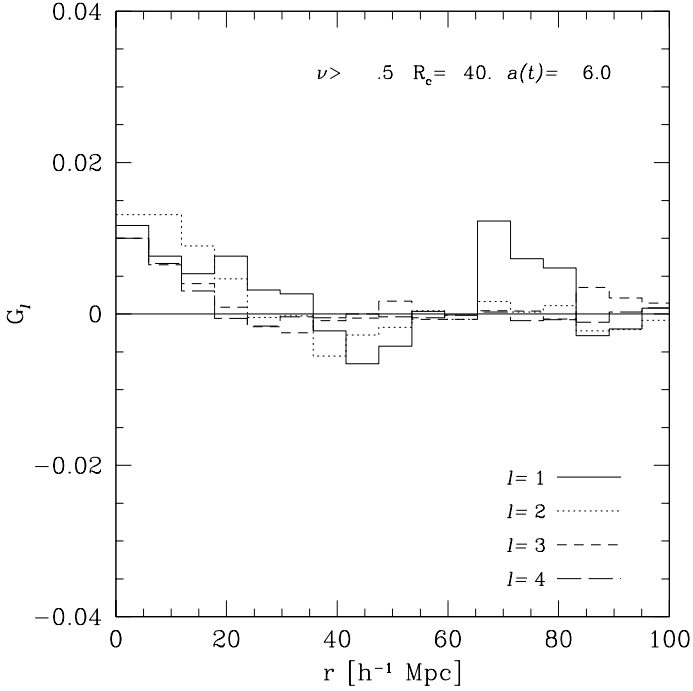,height=4.in,width=3.8in}
}
\caption{ The bond correlation functions $G_l(r)$, divided by $G_0$, are shown
at $a(t)=6$ and $R_c=40$ for the distribution obtained from 
$\{x\}_{s=0.5}$. The different histograms are for different values of $l$, 
only $G_l$ up to $l=4$ are plotted. The line with a value for $G_l$ 
close to zero is for a random distribution.}
\label{Ga}
\end{figure}

\begin{figure} 
\centerline{
\psfig{file=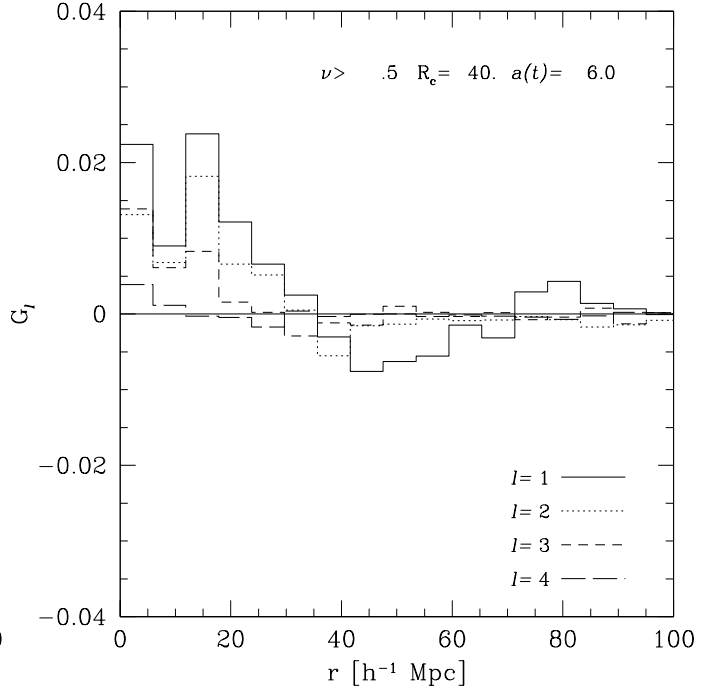,height=4.in,width=3.8in}
}
\caption{ As in Fig. \ref{Ga}, but in this case $G_l(r)$ have been 
evaluated summing in (\ref{eq:GL4}) over all the neighbors of the 
center particles; in the previous figure $G_l(r)$ were calculated 
taking random sub-sample of the neighbor sets.}
\label{Ga1}
\end{figure}

To analyze the clustering shape, more interesting results are obtained 
if the estimator (\ref{eq:GL4}) is applied to point distributions
generated by different $\xs$.  The total operation count  
 of applying Eq. (\ref{eq:GL4}) to a given point distribution scales
as $\propto N_p^4$, thus making it prohibitive to evaluate  
$G_l(r)$ for the whole set $\{x\}_s$  of the simulation box.
Therefore one has to resort to compute $G_l$ taking as centers $i$ in
Eq. (\ref{eq:GL4}) a random subsample of $\{x\}_s$. The summations \\over 
the particles $k$ at a distance $r$ from $i$ and the neighbor particles of the 
chosen centers ($\sum_j \sum_p$) are
instead run over the whole set $\{x\}_s$, although a random dilution of
the neighbor sets was performed in many cases in order to reduce the 
total computational cost. For a given particle $i$ taken as a center the
mean number $\bar N_i$ of neighbors $j$ within a distance $R_c$ from 
$i$ ranges from $\simeq10^2$ to $\simeq10^4$, according to the chosen cut-off 
radius $R_c$, threshold level $\nu$ and expansion factor $a(t)$.

However the use of a random subset requires some care 
with respect to the previous procedures used to compute the coefficients
$Q_l$. It has been found that the function $G_l(r)$ is subject, 
in some cases, to
large sample-to-sample variations when different random subsets
are chosen from a given $\{x\}_s$ to evaluate the corresponding $G_l$.
This shows that $G_l(r)$ is a sensitive measure of the clustering patterns
generated by a point distribution.

\begin{figure} 
\centerline{
\psfig{file=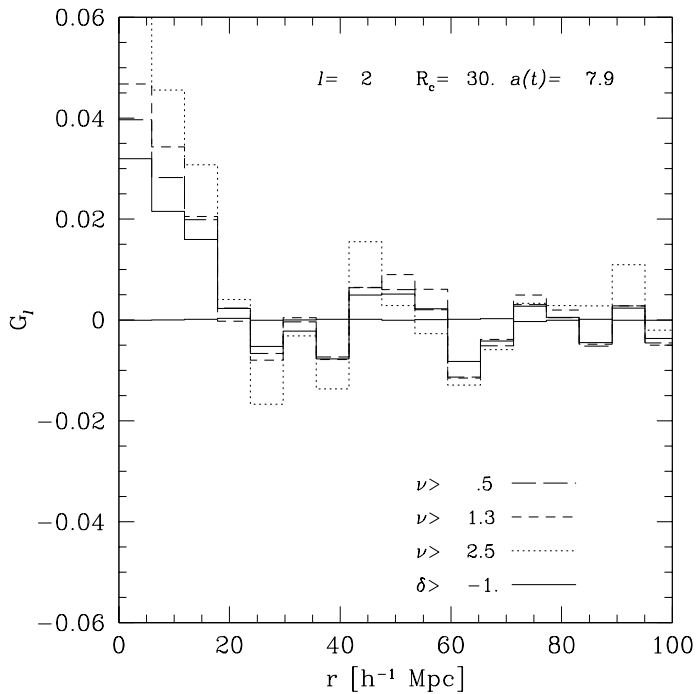,height=4.in,width=3.8in}
}
\caption{ The functions $G_l(r)$ plotted here have been obtained from
different point distributions and the same values of $l,R_c$ and $a(t)$ 
(given in the panel). The distributions are those given by the simulation 
particle subsets $\{x\}_{s}$, with $s=0.5,1.3,2.5$, and  $\{x\}_{s=-1}$.}
\label{Gb}
\end{figure}

One of the consequences of this result is that the calculated $G_l(r)$
analyze in this case the clustering morphology of the point distribution as 
it is
measured according to the selected random sub-sample, 
rather than that of the whole $\{x\}_s$.
In order to consistently compare $G_l(r)$ for \\different choices 
of several parameters, the same random subset has been chosen at different 
times and for different \\$R_c$.
Furthermore $G_l(r)$ for point distributions obtained for
different thresholds have been calculated with random subsamples 
subject to the constraint that for a given $\xs$ the random subset
was part of the random sample corresponding to a lower threshold, 
and so on down to the lowest level with $s=-1$.
The application of these constraints leaves about $8,000$ particles
in the simulation volume, which have been taken as center 
particles when the $G_l(r) $ were calculated.
This procedure allows a consistent analysis to be made of the clustering 
morphologies
detected by the $G_l(r)$ at different epochs and different values of the 
cut-off parameter $R_c$, angular
scale $l$ and threshold $\nu$.
The results obtained are shown in the following plots.
Hereafter $G_l(r)$ is the estimator (\ref{eq:GL4}), applied to point 
distributions, divided by $G_0(r)$.
Because of the large number of cases spanned  in the parameter space, only 
a few of them have been plotted, the main features seen are quite general.

In Fig. \ref{Ga} $G_l(r)$ is given as a function of $r$ for different
values of $l$, up to $l=4$. The values for the other parameters are given 
in the panel. The correlation between angular distributions for
different bonds decreases with increasing $r$ and as smaller scales are probed,
 but with some exceptions. In this case $G_l$, for $l=1$, has a 
significant correlation above the noise extending for a wide range of
spatial distances. In other cases 
a similar trend has been found also for $l=2$. $G_l(r)$ is a measure of 
the degree of
coherence in the clustering patterns. For the example shown in the 
figure the result is a web of large-scale structures, in the examined 
point distribution, which is detected by the estimator.
These features are detected at the lowest order only ($l\leq 2$), since
their distribution in the bond spheres is limited to large angles.
For $r \rightarrow0$ the correlation strength increases because the
two neighbor spheres overlap and partially trace the same structures.
The continuum limit (\ref{eq:Gc}) is reached in the large sample
regime $\bar N_i \rightarrow \infty$. For a finite number of points 
$G_l(r)$ measures the anisotropies of the distribution which samples the 
density field.

A theoretical estimate of how $G_l(r)$ are affected by noise terms is
difficult, therefore an indirect check has been performed for the examples
showed in the plots evaluating $G_l(r)$ in correspondence of a random 
distribution. For each of the considered cases this was obtained by 
considering tha same set of $M_p$ pairs used in Eq. (\ref{eq:GL4}) 
to evaluate $G_l(r)$, but with a random distribution of $N_r=10^3$ 
neighbors within each of the two spheres of radius $R_c$ centered at 
the pair coordinates. The results show that $G_l \simeq 0$ for a 
random distribution of neighbors, so that the measured $G_l(r)$ 
are well above the noise.

For the example showed in Fig. \ref{Ga} $G_l(r)$ have been 
computed using random sub-samples of the $\bar N_i\simeq 10^4$ 
neighbors of the center particles $i$. 
In this case the fraction $f$ of selected neighbors was $f \approx 1/4$. 
Finite sample effects on the measured 
$G_l(r)$ can be estimated by repeating the evaluation of $G_l(r)$ for 
the case of Fig. \ref{Ga}, but this time running the summations in 
(\ref{eq:GL4}) over all the neighbors particles.
The results are showed in Fig. \ref{Ga1}, a comparison with the
previous figure shows that the $G_l(r)$ profiles are qualitatively similar,
suggesting that measurements of the clustering pattern by $G_l(r)$ are
not affected by a dilution of the neighbor sets. Empirically it 
has been found that for the examples showed here this is valid for 
$ f\bar N_i \simgt 10^2$.

$G_l(r)$ is also sensitive to different degrees of clustering, as shown
 in Fig. \ref{Gb}. In this case the $G_l$ have been computed for
different $\xs$ with the other parameters being kept constant (see the panel).

\begin{figure} 
\centerline{
\psfig{file=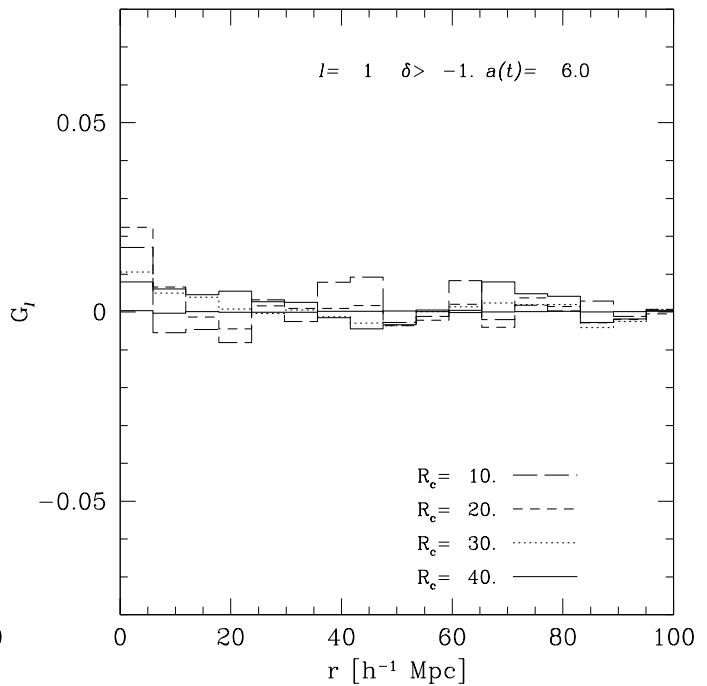,height=4.in,width=3.8in}
}
\caption{ The $G_l(r)$ are shown for different values of the cut-off radius:
$R_c=10,20,30,40h^{-1}Mpc$. The  values of the other parameters are given 
in the panel.} 
\label{Gc}
\end{figure}

It is important to stress that this result was obtained because of the way
in which the random subsets were chosen; if different subsets
had been used for different $\xs$, then sample-to-sample
variations would have dominated over the differences in $G_l$ for
 different thresholds.
A generic feature of $G_l$ is a decrease in the signal as $R_c$ is increased 
(Fig. \ref{Gc}).
This is expected, since in going to higher cut-off radii, the point 
distribution in the bond sphere approaches  isotropy. 

As previously stressed $G_l(r)$ has been found to be very sensitive 
to sample variations, the size of the fluctuations can be checked in 
Fig. \ref{Gd}, where $G_l(r)$ for a specified set of parameters
 has been computed for the different $\xs$ of the numerical ensemble.
Because of the computational cost
of  evaluating $G_l(r)$ I have calculated these functions for the whole 
simulation ensemble only for $a(t)=6$.
The plot shows clearly how fluctuations in the ensemble, in this case,
are of the same order as the variations in $G_l$ shown in the previous plots 
when the input parameters were changed. 

\begin{figure} 
\centerline{
\psfig{file=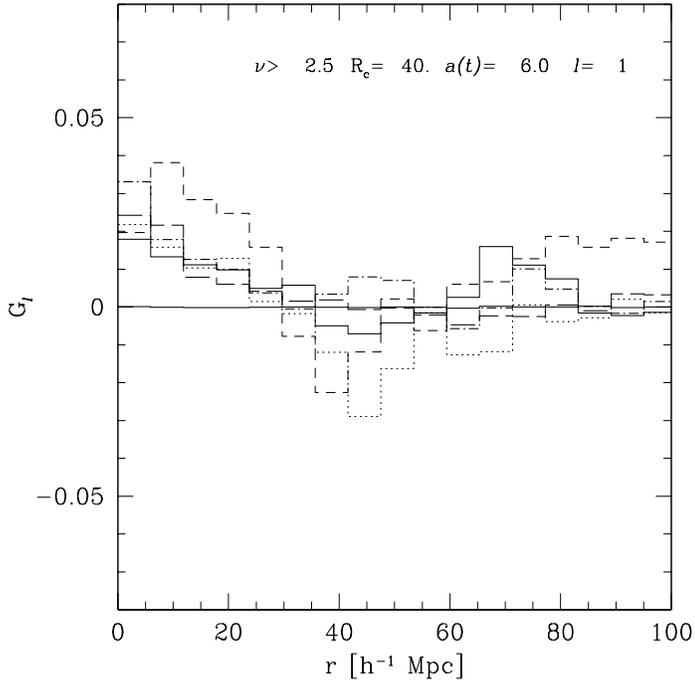,height=4.in,width=3.8in}
}
\caption{ The correlation  functions $G_l(r)$ are plotted here  for
$l=1,R_c=40,a(t)=6$ and point distributions obtained from 
$\{x\}_{s=2.5}$. 
Different histograms refer to 
$\{x\}_{s}$  from simulations with different  
random realizations of the ensemble.}
\label{Gd}
\end{figure}

In one simulation large-scale correlations are clearly detected for 
$r\simgt60 h^{-1} Mpc$.
These results are in accordance with those of Doroshkevich, Fong \& Makarova 
(1998), who have analyzed the evolution of large-scale structure in standard
CDM $N$-body simulations. 
Their analysis of sheets and filaments in the particle distribution 
is based upon the 'core sampling' method (Buryak, Doroshkevich \& Fong 1994).
A proper comparison is difficult because of the different statistical methods 
employed. The results obtained by Doroshkevich, Fong \& Makarova (1998) reveal
a richness of structures on scales of $\sim10 -100h^{-1} Mpc$, in agreement 
with the correlation range measured here by the functions $G_l$.

In Fig. \ref{Ge} the functions $G_l(r)$ are shown, for different 
realizations, for the same example as in Fig. \ref{Gd}, but in this case
for $\{x\}_{s=0.5}$ instead of $\{x\}_{s=2.5}$. The scatter in the ensemble is
clearly reduced; this is a consequence of the lower threshold considered.
For a point distribution approaching that of the whole mass in the 
simulation volume then $G_l(r)$ has a larger contribution from
particles which are not part of filaments or of other large-scale structures.
The signal is thus reduced as also are the variations between different random
realizations.
This is valid for cut-off radii much larger than the non-linear scale.
If $R_c=10h^{-1}Mpc$ then the functions $G_l$ would have shown no sign
of correlations, with strong fluctuations around zero independently
of the separation distance $r$.
The scattering in the ensemble is also reduced when smaller angular scales
are probed (Fig. \ref{Gf}).

\begin{figure} 
\centerline{
\psfig{file=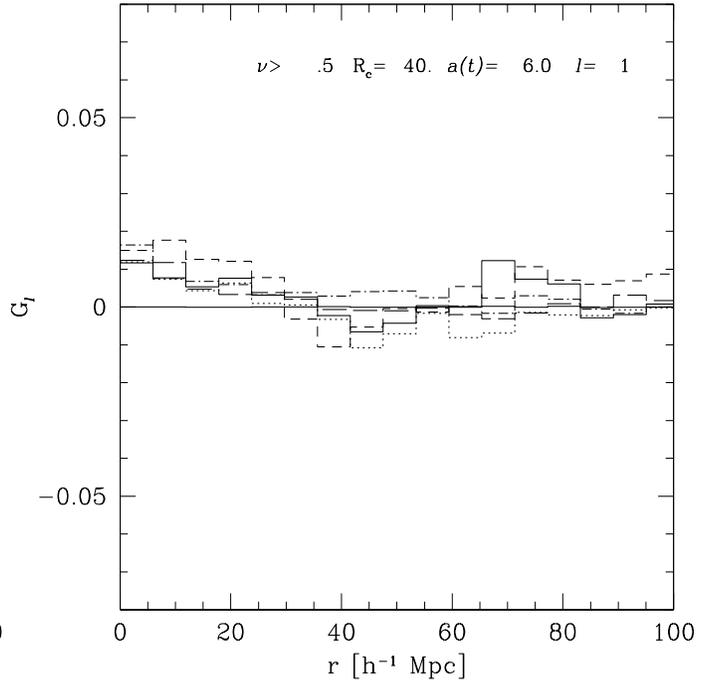,height=4.in,width=3.8in}
}
\caption{ The same as in Fig. \ref{Gd}, but for distributions obtained from
$\{x\}_{s=0.5}$.}
\label{Ge}
\end{figure}

To summarize, the observed results have shown that the statistical method
defined by the functions $Q_l$ and $G_l(r)$ can be used to analyze the
clustering morphology produced by gravitational clustering in a 
quantitative way.
The method was originally devised in a completely different field for 
studying anisotropic structures and its application to cosmological clustering
 shows that, to some extent, it overlaps with previous analyses.
The usefulness of the method has been tested by applying it to a set of 
cosmological N-body simulations with a CDM power spectrum.
An application of the estimators to selected numerical outputs 
shows that the statistic is clearly able to discriminate between 
particle populations with different degrees of clustering.
Several shape statistics have been introduced to quantify the 
presence of filaments or sheet-like structures in the 
clustering network. The function $G_l(r)$ used here describes
anisotropies in the clustering distribution by measuring the degree of 
correlation between the angular densities as seen from two different 
observers separated by $r$.
$G_l(r)$ can then be considered a statistical measure of clustering
patterns, with different scales probed by varying the input
parameters $l$ and $R_c$. This provides an alternative to other 
methods proposed so far 
to quantify the clustering morphology.
\begin{figure} 
\centerline{
\psfig{file=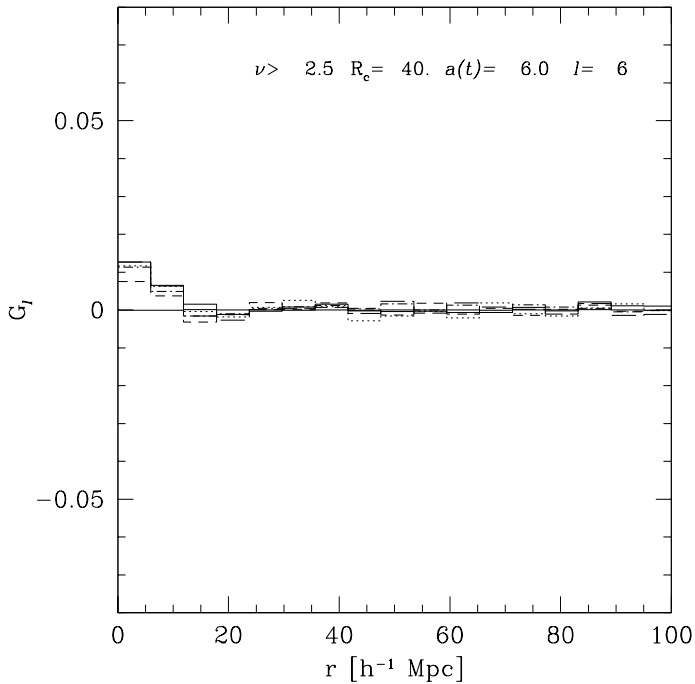,height=4.in,width=3.8in}
}
\caption{ As in Fig. \ref{Gd}, but for $l=6$.}
\label{Gf}
\end{figure}

With large redshift surveys becoming available in the next few years, the
proposed statistical method appears as a promising tool for analyzing
patterns in the galaxy distribution.
In practice real galaxy catalogs have an incomplete sky coverage which 
implies, for the observed harmonic coefficients, non-zero mean values 
and a coupling to different modes of the all-sky coefficients. This
occurs because of the  tensor window function of the angular mask arising from 
partial sky coverage. The required formalism  has already been applied in 
the literature to angular catalogs (Scharf et al. 1992; Scharf et al. 1993) 
and its application to the $Q_l$ coefficient is straightforward. 
Further complications arising from the radial selection function of 
the catalog and redshift space distortions have also been considered  
(Scharf et al. 1993; Ballinger, Heavens \& Taylor 1995).
For the functions $G_l(r)$ the analysis is
much more cumbersome and is beyond the scope
of this paper, which is intended to introduce the method and to illustrate
its features by applications to numerical simulations.

An issue which has not been considered is whether this statistical tool can 
successfully
be used to check the consistency of clustering data with cosmic models
of structure formation.
The analysis performed reveals (Fig. \ref{Gd}) that any method which 
quantifies in a sensible way the presence of filaments or  planar
structures in the cosmic web is also prone to cosmic variance.
When different scales or levels of bias are considered, then the
fluctuations are reduced (Figs. \ref{Ge} \& \ref{Gf}).
The results obtained suggest that an application of the estimator 
$G_l(r)$ to real data will require different scales to be probed, 
in order to give useful constraints on cosmic theories.

\noindent
{\bf Acknowledgements}
The author is grateful to the referee (M.Kerscher) for useful
comments which improved the article.

\end{document}